\documentclass{appolb}
\usepackage{epsfig}

\begin{document}
\title{Analysis of transverse momentum distributions observed at RHIC by a stochastic model
       in hyperbolic space%
\thanks{Presented at XXXIII International Symposium on Multiparticle dynamics, September 5-11, 2003, Krakow, Poland}%
}
\author{ N. Suzuki
\address{Department of Comprehensive Management, Matsumoto University, Matsumoto, Japan}
\and
 M. Biyajima
\address{Department of Physics, Shinshu University, Matsumoto, Japan }
}
\maketitle
\begin{abstract}
To describe large momentum distributions of charged hadrons observed at RHIC, a diffusion equation in the three dimensional hyperbolic space is introduced.
\end{abstract}
  
\section{Introduction}
 We are much intersted in events with high multiplicity at RHIC, where thousands of particles are produced per event.  To describe such many particle system, a sort of collective approach will be useful. 
One-particle rapidity or pseudo-rapidity distributions observed at RHIC 
are well described by the Ornstein-Uhlenbeck process with two or three sources~\cite{biya02}.
In this paper, we would extend our approach to include an expansion for transverse direction.  We can derive the formula same with that proposed by Voloshin~\cite{volo97} from the three-dimensional Ornstein-Uhlenbeck process using the variables, longitudinal rapidity and two transverse momenta. The fundamental solution is gaussian in these variables. 
Even if the formula is integrated over the azimuthal angle in the transverse momentum plane, the distribution is only slightly modified from the gaussian. 

However, observed $p_t$ distributions at RHIC have long tails compared with an exponential distribution, and cannot be described by the formula. Therefore relativistic approach, in other words, a stochastic model in the three dimensional hyperbolic space is adopted for the production processes, and $p_t$ distributions observed at RHIC are analysed by the model. 
\section{Diffusion equation in the three dimensional hyperbolic space}
For simplicity, we consider the diffusion equation in the geodesic polar coordinate system.
In this system, 4 energy-momentum $(E,p_L,{\bf p_T})$ of particle with mass $m$ is specified 
by radial rapidity $\rho$, polar angle $\theta$ and azimuthal angle $\phi$.
Then energy and absolute value of momentum are expressed respectively as
\begin{eqnarray}
    E = m\cosh\rho,   \quad 
    p = \sqrt{p_L^2+{\bf p_T}^2} = m\sinh\rho.   \label{eq.dif1}
\end{eqnarray}
%
%
%
The metric in the geodesic polar coordinate system is written as
\begin{eqnarray*}
  ds^2 =d\rho^2 + \sinh^2\rho \,\,( d\theta^2 + \sin^2\!\theta\, d\phi^2 ).
\end{eqnarray*}

The diffusion equation, 
\begin{eqnarray}
  \frac{\partial f}{\partial t}= \frac{D}{\sinh^2\rho} 
   \left[
      \frac{\partial}{\partial \rho}\left( 
        \sinh^2\!\rho\, \frac{\partial f}{\partial \rho}
      \right)           
  + \frac{1}{\sin\theta}
   \frac{\partial}{\partial \theta}
     \left( \sin\theta \frac{\partial f}{\partial \theta}\right) 
   + \frac{1}{\sin^2\!\theta}
     \frac{\partial^2 f}{\partial \phi^2} \right]   \label{eq.dif3}
\end{eqnarray}
with initial condition
\begin{eqnarray}
    f(\rho,\theta,\phi,t=0)= \frac{\delta(\rho)}
     {\sinh^2\!\rho\, \sin\theta}.     \label{eq.dif4}
\end{eqnarray}
is taken as a model of particle production processes: After a collision of nuclei, particles are produced at the origin of rapidity space expressed by Eq.(\ref{eq.dif4}). Then those particles propagate according to the diffusion equation (\ref{eq.dif3}).  In the course of the space-time development, energy is supplied from the leading particle system (composed of collided nuclei) to the produced  particle system. Number density of particles becomes lower and at some (critical) density particles become free. 

Approximate solution of Eq.(\ref{eq.dif3}) for $Dt<<1$ is given as~\cite{molc75}  
\begin{eqnarray}
  f(\rho,t)&=& C \frac{\rho}{\sinh\rho} 
     \exp\left[-\frac{\rho^2}{2\sigma(t)^2} \right], \nonumber \\
   \sigma(t)^2 &=& 2Dt.   \label{eq.dif5}
\end{eqnarray}
Equation (\ref{eq.dif5}) is the solution of the radial symmetric diffusion equation~\cite{karp59},
\begin{eqnarray}
  \frac{\partial f}{\partial t}&=& \frac{D}{\sinh^2\!\rho}\, 
      \frac{\partial}{\partial \rho}\left( 
        \sinh^2\!\rho\, \frac{\partial f}{\partial \rho} 
      \right)           \label{eq.dif6}
\end{eqnarray}
if $C=(4\pi Dt)^{-3/2} {\rm e}^{-Dt}$.
In the geodesic cylindrical coordinate system, 4 energy-momentum
 is expressed by longitudinal rapidity $y$, transverse rapidity $\xi$ and azimuthal angle $\phi$.  
 Energy is given as 
\begin{eqnarray}
    E = m\cosh y \cosh\xi =m_{\rm T}\cosh y. \label{eq.dif7}
\end{eqnarray}
where $m_{\rm T}=\sqrt{p_{\rm T}^2+m^2}$. 
From Eqs.(\ref{eq.dif1}) and (\ref{eq.dif7}), we have an identity,
 \begin{eqnarray}
   \cosh \rho = \cosh y \cosh\xi.  \label{eq.dif8}
 \end{eqnarray}
%
%
Radial rapidity coincides with transverse rapidity, $\rho=\xi$, when $y=0$.

Therefore, we can analyse transverse momentum (rapidity) distributions at fixed longitudinal rapidity ($y=0, 2.2, \cdots$), using Eqs.(\ref{eq.dif5}) and (\ref{eq.dif8}) 
with parameters, $C$ and $\sigma(t)^2$
\section{Analysis of $p_t$ distributions observed at RHIC}

Transverse momemtum distributions of charged hadrons, $h^+$ and $h^-$, observed by the STAR ~\cite{adam03} and BRAHMS~\cite{arse03} collaborations are analysed.   Data are taken at fixed pseudo-rapidity $\eta$. Particles are not identified in these experiments. Therefore, we use mass $m$ as a parameter in addition to $C$ and $\sigma^2(t)$.  To estimate the values of parameters, only the statistical error is used.

The results at $\eta=0$ by the STAR collaboration are shown in Fig.1 and Table 1. Estimated value of $m$ decreases as the centrality cut increases, where as variance $\sigma^2(t)$ decreases.

The results at $\eta=0$ by the BRAHMS collaboration are shown in Fig.2 and Table 2. Estimated value of m increases as centrality cut increases. However the value is somewhat higher than that by the STAR collaboration at the same centrality cut.  Variance $\sigma^2(t)$ decreases as the centrality cut increases. In this case the value is smaller than that by the STAR collaboration.

\begin{center}
 \begin{tabular}{ccccc}  \hline
  centrality & $m$ & $C$ & $\sigma(t)^2$ & $\chi^2_{min}/$n.d.f \\ \hline
  00-05\% & 0.57 &  316.7 $\pm$ 5.2 & 0.332$\pm$ 0.001 & 244.9/32\\ \hline
  05-10\% & 0.56 &  266.5 $\pm$ 4.7 & 0.334$\pm$ 0.001 & 129.6/32\\ \hline
  10-20\% & 0.53 &  202.8 $\pm$ 3.1 & 0.354$\pm$ 0.001 & 144.9/32\\ \hline
  20-30\% & 0.52 &  143.5 $\pm$ 2.5 & 0.378$\pm$ 0.001 &  97.94/32\\ \hline
  30-40\% & 0.50 &  101.2 $\pm$ 2.5 & 0.360$\pm$ 0.001 &  98.3/32\\ \hline
  40-60\% & 0.44 &   62.7 $\pm$ 1.1 & 0.402$\pm$ 0.001 & 52.1/32\\ \hline
  60-80\% & 0.37 &   25.6 $\pm$ 0.5 & 0.444$\pm$ 0.001 & 26.6/32\\ \hline
  min-bias& 0.32 &   3.03 $\pm$0.08 & 0.457$\pm$ 0.004 & 16.1/29\\ \hline
\end{tabular}

Table 1.  Estimated parameters of $p_t$ distributions of $\frac{h^++h^-}{2}$ 
at $\eta=0$ in Au+Au collisions at $\sqrt{s_{\rm NN}}$=200 GeV~\cite{adam03}.
\end{center}

 \begin{figure}
  \begin{center}
  \epsfig{file=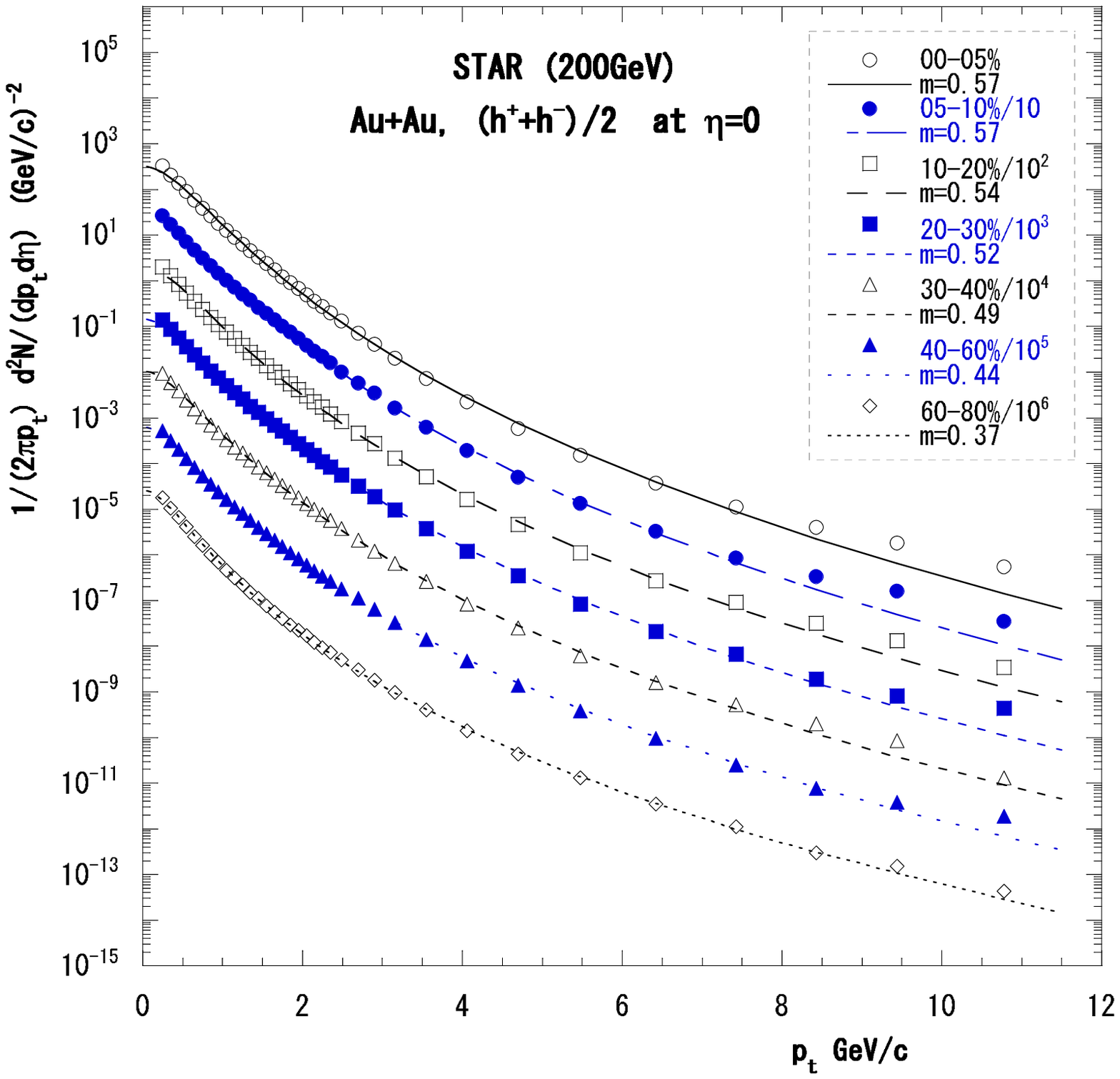, scale=0.40, bb=10 180  500  640, clip }
  \caption{$p_t$ distributions of $\frac{h^++h^-}{2}$ at $\eta=0$
  in Au+Au collisions ~\cite{adam03} }
  \end{center}
 \end{figure}
%


%
\begin{figure}
\begin{minipage}{0.45\linewidth}
\epsfig{file=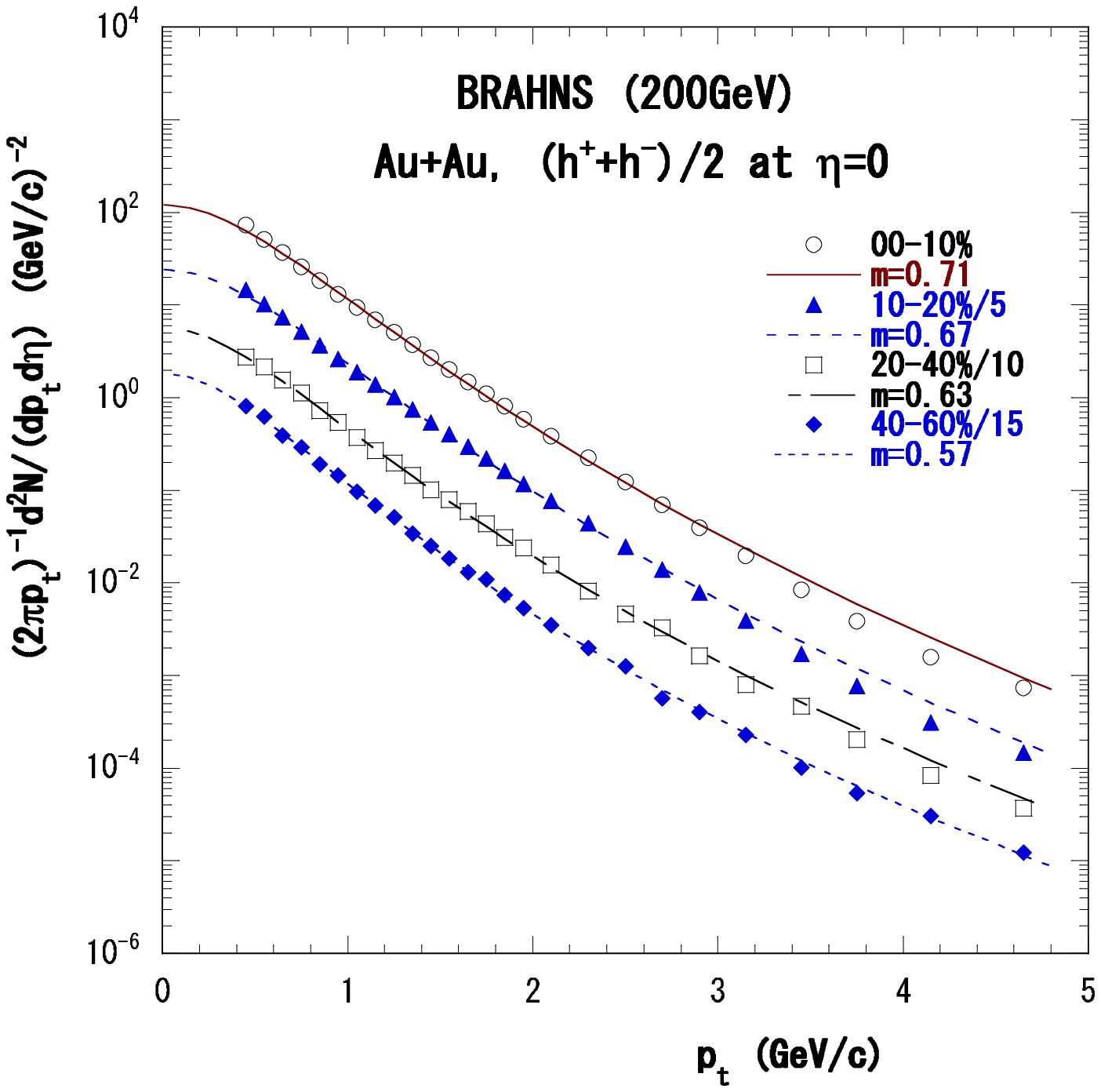, scale=0.4, bb=55 200  500  600, clip }
\caption{$p_t$ distributions of $\frac{h^++h^-}{2}$ at  $\eta=0$ in Au+Au collisions~\cite{arse03}}
 \end{minipage}
 \hspace{5mm}
 \begin{minipage}{0.45\linewidth}
   \epsfig{file=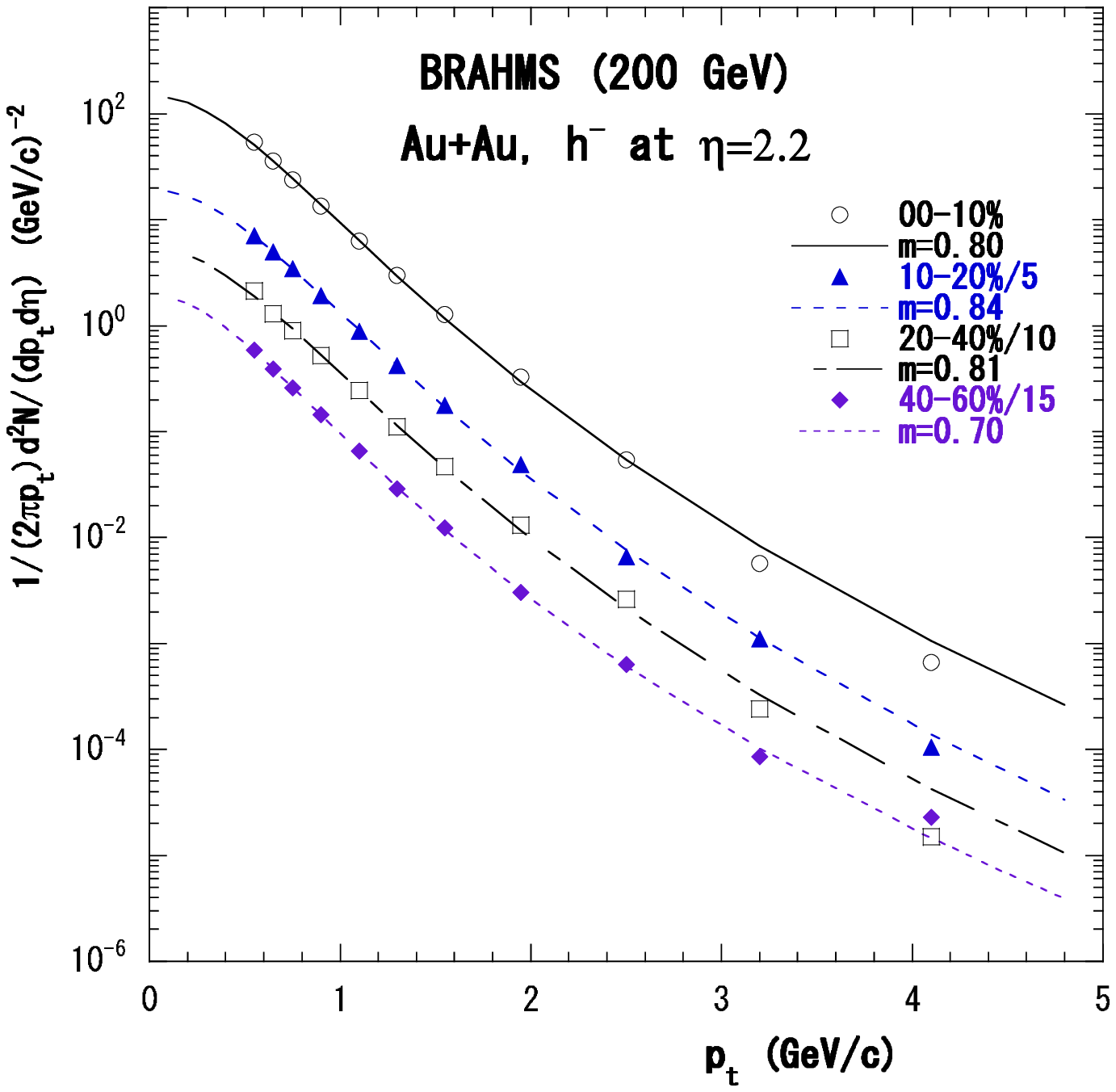, scale=0.4, bb=60 205  500  615, clip }
   \caption{$p_t$ distributions of $h^-$ at  $\eta=2.2$ in Au+Au collisions~\cite{arse03}}
 \end{minipage}
 \end{figure}

\vspace{-1cm}
%
\begin{center}
\begin{tabular}{ccccc}  \hline
  centrality & $m$ &  $C$  & $\sigma(t)^2$   & $\chi^2_{min}$ /n.d.f\\ \hline
  00-10\% & 0.71  & 121.0 $\pm$ 0.7 & 0.307$\pm$ 0.001 & 351.2/23 \\ \hline
  10-20\% & 0.67  &  88.7 $\pm$ 0.6 & 0.325$\pm$ 0.001 & 233.5/23 \\ \hline
  20-40\% & 0.63  &  56.9 $\pm$ 0.7 & 0.340$\pm$ 0.001 & 97.8/23 \\ \hline
  40-60\% & 0.57  &  27.5 $\pm$ 0.5 & 0.358$\pm$ 0.002 & 28.7/23 \\ \hline
\end{tabular}

  Table 2. 
  Estimated parameters of $p_t$ distributions of $\frac{h^++h^-}{2}$ at $\eta=0$
in Au+Au collisions  at $\sqrt{s_{\rm NN}}$=200 GeV by the BRAHMS collaboration~\cite{arse03}.
\end{center}

The results at $\eta=2.2$ by the BRAHMS collaboration are shown in Fig.3 and Table 3. 
Data are well described by Eq.(\ref{eq.dif5}). In this analysis, we assume y=2.2, 
and use Eq.(\ref{eq.dif8}). 

%
\begin{center}
 \begin{tabular}{ccccc}  \hline
  centrality & $m$ &  $C$  & $\sigma(t)^2$  & $\chi^2_{min}$/n.d.f \\ \hline
  00-10\% & 0.80  &  52286 $\pm$ 1575 & 0.470$\pm$ 0.002 & 93.7/8 \\ \hline
  10-20\% & 0.84  &  41669 $\pm$ 1561 & 0.452$\pm$ 0.002 & 30.5/8 \\ \hline
  20-40\% & 0.81  &  19364 $\pm$  862 & 0.469$\pm$ 0.003 & 68.9/8 \\ \hline
  40-60\% & 0.70  &   6070 $\pm$  418 & 0.524$\pm$ 0.005 &  3.1/8 \\ \hline
\end{tabular}
%

 Table 3.  
 Estimated parameters of $p_t$ distributions of $h^-$ at $\eta=2.2$
in Au+Au collisions at 200 GeV by the BRAHMS collaboration~\cite{arse03}.
\end{center}

\section{Summary and discussions}

In order to analyse large $p_t$ distributions of charged hadrons observed at RHIC, a stochastic process in the three dimensional hyperbolic space is taken as a model of particle production process. The solution is gaussian-like in radial rapidity.

Transverse momentum distributions at $\eta=0$ observed by the STAR collaboration and those at $\eta=0$ and $ 2.2$ by the BRAHMS collaboration are well described by the formula, if the value of 
mass contained in rapidity variable is used as a parameter.

\vspace{5mm}
{\bf Acknowledgement}

Authors thank Calderon for his kind correspondence.

\end{document}